\documentclass[5p,twocolumn,preprint]{elsarticle}

\usepackage{amsmath}
\usepackage{amssymb}
\usepackage{graphicx}
\usepackage{upgreek}

\newcommand{\be}{\begin{equation}}
\newcommand{\ee}{\end{equation}}
\newcommand{\bs}{\begin{subequations}}
\newcommand{\es}{\end{subequations}}

\newcommand{\rmd}{\mathrm{d}}
\newcommand{\rmi}{\mathrm{i}}

\newcommand{\bu}{\mathbf{U}}
\newcommand{\bO}{{\boldsymbol\Omega}}

\newcommand{\bey}{\mathbf{e}_y}
\newcommand{\bez}{\mathbf{e}_z}
\newcommand{\bk}{\mathbf{k}}
\newcommand{\br}{\mathbf{r}}
\newcommand{\hu}{\hat{u}}
\newcommand{\hv}{\hat{v}}
\newcommand{\hw}{\hat{w}}
\newcommand{\hp}{\hat{p}}

\begin{document}

\begin{frontmatter}

\title{Oblique waves on a vertically sheared current are rotational}

\author[ntnu]{Simen {\AA}. Ellingsen\corref{cor}\fnref{fax}}
\ead{simen.a.ellingsen@ntnu.no}

\cortext[cor]{Corresponding author.}
\fntext[fax]{Fax number: +47 73593491. Telephone: +47 73593554.}
\address[ntnu]{Department of Energy and Process Engineering, Norwegian University of Science and Technology, N-7491 Trondheim, Norway}

\begin{abstract}
  In the study of surface waves in the presence of a shear current, a useful and much studied model is that in which the shear flow has constant vorticity. Recently it was shown by Constantin [Eur.\ J.~Mech.\ B/Fluids 30 (2011) 12--16] that a flow of constant vorticity can only permit waves travelling exactly upstream or downstream, but not at oblique angles to the current, and several proofs to the same effect have appeared thereafter.  Physical waves cannot possibly adhere to such a restriction, however. We resolve the paradox by showing that an oblique plane wave propagating atop a current of constant vorticity according to the linearized Euler equation carries with it an undulating perturbation of the vorticity field, hence is not prohibited by the Constantin theorem since vorticity is not constant. The perturbation of the vorticity field is readily interpreted in a Lagrangian perspective as the wave motion gently shifting and twisting the vortex lines as the wave passes. In the special case of upstream or downstream propagation, the wave advection of vortex lines does not affect the Eulerian vorticity field, in accordance with the theorem. We conclude that the study of oblique waves on shear currents requires a formalism allowing undulating perturbations of the vorticity field, and the constant vorticity model is helpful only in certain 2D systems.
\end{abstract}


\end{frontmatter}

\section{Introduction}

The interaction of surface waves and currents has been a topic of interest for a long time. In the presence of depth-dependent current, the nature of surface waves can change perceptibly, a problem of great technological relevance in areas where currents with near-surface vorticity are often present, such as the near-shore region and river deltas, and in the presence of currents generated by wind or tides \cite{peregrine76}. For example, the vorticity of the tidal current in the Columbia River mouth was reported at around $0.4$s$^{-1}$ in the top $5$ metres of the water column, enough to significantly affect the dispersion of gravity waves of wavelengths up to tens of metres \cite{dong12}. Waves interacting with currents could be one of the key mechanisms for the generation of giant waves \cite{kharif03,toffoli15}, and have been considered lately both analytically and numerically for vertically sheared currents where the current itself has constant vorticity \cite{nwogu09,thomas12}. Wave-current interactions have also been studied in the context of wave resistance for insect biolocomotion \cite{benzaquen12,bush06}. For two-dimensional systems of waves and shear currents, a sizeable literature exists \cite[see, e.g.,][and references therein]{peregrine76,ellingsen14c,buhler09}, but with very few exceptions \cite[e.g.,][]{craik68,johnson90,mchugh94} wave propagation other than directly with or against the current have not been studied. The shear current of constant vorticity with a free surface has recently also attracted much interest in the mathematical community, e.g.\ \cite{ehrnstrom08,wahlen09,constantin11a,constantin11b} and further references therein.

Recently Constantin \cite{constantin11} proved that when the vorticity is constant for a shear with a free surface, wave propagation must be aligned either exactly upstream or exactly downstream, i.e., the flow must be effectively two-dimensional. His work furthers that of Constantin \& Kartashova \cite{constantin09} and additional proofs of similar results have followed \cite{stuhlmeier12, wahlen14}. While the mathematical argument, briefly recounted below, is indisputable, the result proved in these references appears to run counter not only to physical intuition, but also appears to make the use of vectorial Fourier analysis for linear surface waves in 3D illegal when a shear flow is present, since oblique Fourier components would be forbidden. 

On the other hand, recent times have seen progress made on the fully 3D problem of linear surface waves on top of a shear current of constant vorticity, and explicit solutions to the linearized Euler equations for 3D surface waves on such a background flow have been found for initial value problems \cite{ellingsen14b,li15b}, ship waves \cite{ellingsen14a,li15c}, and waves from a submerged oscillating source \cite{ellingsen15b}. In all cases the solutions rely on the ability to express wave associated surface elevation and velocity components as a 2D Fourier integral in the horizontal plane, whose kernel functions can be understood as plane waves propagating in \emph{all} horizontal directions. In fact, with only directly upstream and downstream wave components, neither ship wave patterns nor ring waves from an initial disturbance or a point source are possible.

A paradox emerges, therefore, because one and the same equation of motion, the Euler equation, on the one hand necessitates Constantin's theorem of two-dimensionality, and on the other hand permits solutions which are patently three-dimensional. In the following we resolve this paradox. We show that the Euler equation permits periodic plane wave solutions propagating at an angle with the underlying shear flow, but that these waves are required to carry an associated vorticity field except in the special case where wave propagation is aligned with the shear flow.  Constantin's theorem rests on an assumption that the vorticity is constant everywhere, hence there is no contradiction between the theorem and the recent 3D solutions to the Euler equations in the presence of a shear flow. These skew waves are thus, in an Eulerian sense, rotational, although a Lagrangian perspective shows how the rotationality is but a slight redistribution of the vorticity of the underlying current. We show for the oblique linear plane wave, how the vorticity perturbation may be interpreted as the wave motion gently shifting, twisting, stretching and contracting the vortex lines as it travels past.

The basic plane wave solutions for an oblique wave on a linear-profile shear flow are presented, being the building bricks from which above cited results for ship waves, ring waves and the oscillating source were constructed.  
We propose that the theorems, which prove the non-existence of irrotational oblique waves on a shear flow with constant and horizontal vorticity, should be understood physically in a positive sense: skew waves on such a flow must always carry a corresponding vorticity perturbation, i.e., they are themselves, in an Eulerian sense, rotational. The assumption of constant vorticity, while a tempting simplification, is not helpful for the study of three-dimensional waves on shearing currents, and may be employed in some 2D systems only.

\section{Two--dimensionality of constant vorticy waves}

We begin by briefly recounting the result proved by Constantin \cite{constantin11}.
Let us assume that a wave motion appears as a perturbation of a shear flow whose vorticity is constant in time and space and horizontally oriented. 
The flow can have finite or infinite depth, and is presumed to be inviscid and incompressible, hence the velocity field $\bu$ is governed by the Euler equation of motion (a dot denotes partial derivative w.r.t.\ time),
\be
  \dot{\bu} + (\bu\cdot\nabla)\bu = - (1/\rho)\nabla P -g\bez
\ee
and the continuity equation $\nabla\cdot\bu=0$. Here the pressure field is $P$, $g$ is the gravitational acceleration.
Applying the curl operator yields the vorticity equation,
\be\label{vort}
  \dot{\bO} + (\bu\cdot\nabla)\bO = (\bO\cdot \nabla)\bu.
\ee
where the vorticity is related to $\bu$ by $\bO=\nabla\times\bu$. 

The key assumption now made is that the velocity field has constant vorticity $\bO$ in time and space, not only the original shear flow, but the perturbation due to the presence of waves as well. In this case Eq.~\eqref{vort} reduces to 
\be
  (\bO\cdot \nabla)\bu=0,
\ee
i.e., the velocity field can have no variation in the direction parallel to $\bO$. But a wave train propagating in a general direction $\bk$ in the $xy$ plane must, regardless of its shape, be associated with a velocity field which varies along its direction of propagation, hence only waves propagating either directly upstream or downstream with respect to the shear flow can exist. 

In particular, the above theorem implies that a plane wave of constant vorticity, which has the form 
\[
  \zeta(\br,t) \propto \exp[\rmi\bk\cdot\br - \rmi\omega(\bk)t] 
\]
(the real part is understood to be taken) with $\zeta$ the surface elevation and $\br=(x,y)$, must have $\bk$ pointing either exactly upstream or downstream so that $\bk\cdot\bO=0$. 

For linearized wave theory this seems to disagree with the use of a Fourier description, by which \emph{any} surface deformation can be expressed in such a form, with the appropriate eigenvalue for $\omega(\bk)$, and where contributions from $\bk$ in all directions are required to describe, e.g., a localised initial surface perturbation.

Moreover such a conclusion is in discord with physical intuition. A local initial perturbation of a still water surface is a classical problem considered by Cauchy and Poisson 200 years ago \cite{cauchy1816,poisson1818}, and results in ring waves propagating out in all directions, with wave fronts becoming approximately plane far from the origin. The equivalent system with a uniform (irrotational) current can be found by an appropriate Galilei transformation, and the conclusion remains the same. However, Constantin's theorem seems to indicate that if a constant vorticity is now introduced, \emph{however small}, it would drastically change the surface waves, since only up- and downstream propagation is allowed. 

While there is no doubt about the soundness of the theorem itself, it seems clear that real-life wave systems cannot possibly adhere to it. The paradox is resolved in the following.

\section{Solution of the linearised Euler equation for a skew plane wave}

The theorems of Constantin \cite{constantin11}, Wahl\'{e}n \cite{wahlen14} and others assume the full velocity field including the wave motion to have constant vorticity. In systems where the wave motion can be seen as a perturbation of a constant vorticity shear flow (linear and weakly non-linear system), the assumption implies that the wave motion alone carry a constant vorticity, and since it is the nature of wave motion to vary in time and horizontal space (e.g., periodically) and to decrease with depth, the assumption then realistically means the wave motion alone be irrotational.

To elucidate the situation, let us compare with the solution obtained for an oblique wave with no such restrictions imposed. Consider the linearised Euler equation, for a shear flow which itself has uniform vorticity and is of the form
\be\label{U}
  U_\text{curr}(z) = U_0 + Sz
\ee
where the undisturbed surface is at $z=0$ and the shear flow points along the $x$ direction. The velocity field is 
\be
  \bu = (U_\text{curr}+\hu,\hv,\hw) 
\ee
where $\hu,\hv,\hw$ are perturbations due to the wave field, and $  P = -\rho g z + \hp $ with perturbation $\hp$. We consider solutions to linear order in perturbation quantities.

A plane wave is now presumed to travel upon the shear flow \eqref{U} at an arbitrary angle $\theta$ with the $x$ axis, i.e., the perturbation quantities $\hu,\hv,\hw$ and $\hp$ are all presumed to have the form 
\begin{align}
  [\hu,\hv,\hw,\hp]&(\br,z,t) = 
  [u(z),v(z),w(z),p(z)] e^{\rmi \bk\cdot\br - \rmi \omega t}
\end{align}
where $\br=(x,y)$ and the wave vector is $\bk=(k_x,k_y)=k(\cos\theta,\sin\theta)$. $\theta$ is the angle between wave propagation and shear current. The eigenvalues $\omega(\bk)$ that permit a solution are provided by the free-surface boundary conditions. 
The system is similar to that considered in Ref.~\cite{mchugh94} and \S IV.B.3 of Ref.~\cite{peregrine76}. The Euler and continuity equations become
\bs
\begin{align}
  -\rmi \omega u + \rmi k_xU(z)u + Sw =& -\rmi k_xp/\rho,\\
  -\rmi \omega v + \rmi k_xU(z)v =& -\rmi k_yp/\rho,\\
  -\rmi \omega w + \rmi k_xU(z)w =& - p'/\rho,\\
  \rmi k_x u + \rmi k_y v +w'=& 0
\end{align}
\es
(a prime denotes differentiation w.r.t.\ $z$). Eliminating $u,v,p$ we obtain the Rayleigh equation for this case, $w''-k^2w=0$. For simplicity, assume infinitely deep water (finite water depth, considered, e.g., in Ref.~\cite{ellingsen14b}, does not affect the argument essentially, but clutters the formalism). The solutions are then found as
\bs\label{solution}
\begin{align}
  u(z) =& \left[\rmi \cos\theta +\frac{\rmi S\sin^2\theta}{k_xU(z)-\omega}\right]A(\bk) e^{kz}\\
  v(z) =& \left[\rmi \sin\theta -\frac{\rmi S\cos\theta\sin\theta}{k_xU(z)-\omega}\right]A(\bk) e^{kz}\\
  w(z) =& A(\bk) e^{kz}\\
  p(z)/\rho=& -\frac \rmi k \left[k_xU(z)-\omega - S\cos\theta\right]A(\bk) e^{kz},
\end{align}
\es
where the boundary condition $\lim_{z\to -\infty}w=0$ was employed. The kinetic and dynamic boundary conditions now relate the complex-valued amplitude $A(\bk)$ to the surface elevation, and give the dispersion relation for $\omega(\bk)$. For our present purposes these details are not important, and we simply regard $A(\bk)$ and $\omega(\bk)$ as known quantities which can be found with the procedures of e.g., Refs.~\cite{ellingsen14a,ellingsen14b}. 

Consider now the vorticity vector $\bO$. A little calculation yields
\begin{align}
  \bO =& S\bey+SA(\bk)\sin\theta \left[\frac{\rmi \bk+k\bez}{k_xU(z)-\omega}\right.\notag \\
  &\left.-\frac{\rmi \bk S\cos\theta}{[k_xU(z)-\omega]^2}\right] e^{kz+\rmi \bk\cdot\br - \rmi \omega t}.
  \label{omega}
\end{align}

The solution of Eq.~\eqref{omega} shows two things. Firstly, that a plane wave travelling at an angle with the mean flow is a solution to the Euler equation (provided, of course, an appropriate coefficient $A(\bk)$ can be found to satisfy boundary and initial conditions), and secondly, that this solution must carry a non-constant vorticity except when $\sin\theta=0$. Hence there is no contradiction between this solution and Constantin's theorem, which presumed constant vorticity. It also means that the solutions reported in Refs.~\cite{ellingsen14b,li15b,ellingsen14a,li15c,ellingsen15b} are unaffected by the theorem. 

Note that in our linear theory, the amplitude $A(\bk)$ plays the role of a smallness parameter, and all quantities of order $A^2$ and higher are neglected. We have implicitly assumed $A$ to be the only infinitesimal quantity, $A\sim \epsilon\ll 1$, assuming all other quantities to be order unity compared to relevant time and length scales. Even in a linearised perturbative setting, other solutions are possible if quantities such as $S$  be either very small (e.g., $S\sim \epsilon$) or very large (e.g. $S\sim \epsilon^{-1}$). The former is contained in the above as a special case, and is curious because all terms containing $S$ disappear from $u,v,w,p$ as well as the second and third term of Eq.~\eqref{omega}. Hence, a wave of infinitesimal amplitude can propagate at any angle on on a current without introducing additional vorticity, if the vorticity $S$ is also infinitesimal. This observation is perhaps primarily of mathematical interest. The case $S\sim\epsilon^{-1}$ (or similar) would require further analysis, but does not seem particularly pertinent since large values of $S$ rarly occur near a free surface (unlike a solid boundary where no-slip conditions apply). A number of non-linear oblique travelling wave solutions are also certain to exist, not considered herein. In conclusion, the solutions in Eq.~\eqref{solution} are in no way exhaustive, but a significant group of linear surface wave scenarios can be constructed from them, and as such they are sufficient for illustration purposes in the present context.

\section{Vortex dynamics}

The additional vorticity (proportional to $A(\bk)$) in Eq.~\eqref{omega} is a vector in the plane formed by $\bk$ and the $z$ axis, and whose direction rotates to draw an ellipse during an oscillation period. It is the rotational motion of the wave which gently lifts, pushes, twists and stretches the vortex lines of the shear flow, which would otherwise be uniformly spaced and pointed along the $y$ axis. The total vorticity contributed by the wave from one wavelength (or period) is zero, as it must be due to the circulation theorem which asserts that the number of vortex lines passing through a closed, material curve must be constant.

\begin{figure*}
  \begin{center}
   \includegraphics[width=.7\textwidth]{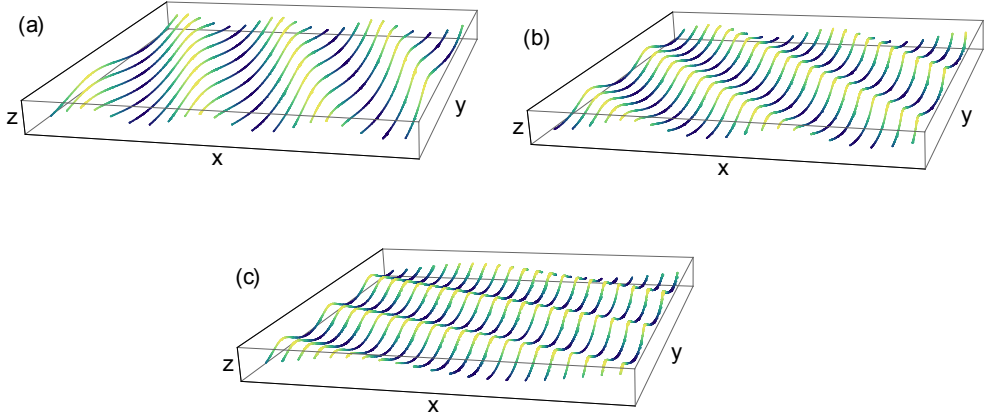}
  \end{center}
  \caption{Vortex lines gently shifted and twisted by a passing wave of small amplitude. Here $A(\bk)\exp(kz_0)=0.3$, $[k_x U(z_0)-\omega]=1$ and $S=0.5$ for $x_0$ in steps of $1$ and $k=1$ (all in arbitrary units). The shear flow is along the $x$ axis, and the angle of propagagation relative to the direction of the shear flow is (top left to bottom) $\pi/8, \pi/4, 3\pi/8$.}
  \label{fig:lines}
\end{figure*}

The additional vorticity field permits a simple interpretation if one takes the Lagrangian view of following the motion of the vortex lines. 
The equation for a vortex line (i.e., a curve which is everywhere tangential to $\bO$) is
\[
 \frac{\rmd x}{\Omega_x} = \frac{\rmd y}{\Omega_y}=\frac{\rmd z}{\Omega_z}
\]
where $\rmd{\boldsymbol s}=(\rmd x,\rmd y,\rmd z)$ points along the vortex line, and $\Omega_x$ is the $x$-component of $\bO$, etc. A vortex line positioned at $(x,z)=(x_0,z_0)$ in the absence of skew waves now obtains the form
\begin{align}
  (x_\text{vl},z_\text{vl})&(y,t) = (x_0,z_0) \notag \\
  &+ \left\{\frac{(\cos\theta,-\rmi)}{k_x U(z_0)-\omega}-\frac{(1,0)S\cos^2\theta}{[k_x U(z_0)-\omega]^2}\right\}\notag \\
  &\times A(\bk) e^{kz_0  +\rmi k_x x_0} e^{\rmi k \sin\theta y- \rmi \omega t} 
\end{align}
where we have kept terms to linear order in $A(\bk)$ as before. The vortex lines are distorted into elliptical helix shapes, a disturbance which is small (for a linear wave) except when $k_xU(z_0) \approx \omega$ when the distortion becomes more violent, particularly in the $xy$ plane. If a depth exists where $k_xU(z_0) = \omega$ this corresponds to a critical layer, to be discussed in the following. The shifting and twisting of the vortex lines is illustrated in Fig.~\ref{fig:lines} for three different propagation angles, where for simplicity $(k_x U(z_0)-\omega)$ are taken to be the same in all cases. In the special case when $\sin\theta=0$, i.e., waves propagating exactly upstream or downstream, the vortex lines remain straight and parallel during their wave-induced motion, and the Eulerian vorticity field remains constant.

\section{Further discussion}

The undulating vorticity in Eq.~\eqref{omega} is notable for being singular when $k_xU(z)-\omega=0$, stemming from the corresponding singularity in $u$ and $v$. 

Firstly, singular velocity compoents in the $\bk$ plane is not dramatic or unphysical. Indeed in wave systems which forces the frequency to take certain values, such as stationary systems (e.g., ship waves \cite{ellingsen14a,li15b}) or periodic systems (such as the oscillating source \cite{ellingsen15b}), the amplitude $A(\bk)$ will have a second singularity at the value of $\bk$ where the enforced frequency equals that from the dispersion relation prescribed by the free surface boundary conditions, in our case \cite{ellingsen14b}
\be
  \omega(\bk) = k_xU_0 - (S/2)\cos\theta \pm \sqrt{gk + (S^2/4)\cos^2\theta}.
\ee
The latter singularity is well known, and treated with standard procedures (see further discussion, e.g., in \S3.9 of \cite{lighthill78}).
These values of $\bk$ nominally produce infinite values for a monochromatic plane wave. However, a real wave source will produce waves with a nonzero bandwith of values for $\bk$, both in magnitude and direction. Now integrating over $\bk$ space (invoking an appropriate radiation condition) the pole singularities in $u$ and $v$ each give a finite contribution to the resulting velocity field. Indeed, only these poles can contribute to the wave field far away from the wave source. Likewise, the pole singularity of $u$ and $v$ at one particular $\bk$ provides velocities of finite value once the appropriate integral is taken. Note, however, that horizontal velocity components can be discontinuous at some depth $z$, corresponding if so to a thin vortex sheet at this depth. See also the discussion in \S IV.B.3 of \cite{peregrine76}. 

Secondly we notice at the depth where $\bO,u$ and $v$ are infinite, the phase velocity $c=\omega/k$, when measured along the direction of the current exactly equals the flow velocity, $U(z)=c(\bk)/\cos\theta$. In the theory of linear waves on a general shear current this is known as the criterion for a critical layer to form. When $U''(z)\neq 0$, the Rayleigh equation for $w(z)$ becomes singular at this depth (see, e.g., \S22 of Ref.~\cite{drazin04}), while in the special case of constant vorticity, $U''(z)=0$, and the Rayleigh equation is regular, but a kind of critical layer nevertheless shows up in the velocity components parallel to the undisturbed surface, provided wave propagation is not exactly upstream or downstream. In inviscid theory this horizontal motion does not affect the surface waves directly, a result which holds also for more general shear profiles (see \S IV.B.3 of Ref.~\cite{peregrine76}).

These topics are discussed much further in Ref.~\cite{ellingsen15b} in the context of a point source. We will not delve further into the formation of critical layer-type phenomena on a shear flow of constant vorticity other than to note that they affect even waves on a constant-vorticity shear profile in 3D flow. Note, however, that critical layer solutions exist in the 2D version of the constant-vorticity wave system as well, as pointed out by Ehrnstr\"{o}m \& Villari \cite{ehrnstrom08} and Wahl\'{e}n \cite{wahlen09}. A concrete example where such solutions appear, is when an oscillating line source is inserted \cite{ellingsen15a}, which is a simple model for a real wave-making device.

\section{Conclusions}

We have considered the linearised Euler equations for a shear flow of uniform vorticity with a free surface, and find that basic plane wave solutions are possible in all propagation directions, not only directly upstream and downstream. When the wave is at skew angle with the shear flow, it is associated with an undulating perturbation of the vorticity field. This resolves what might be considered a paradox presented by recent theorems by Constantin and others which precluded skew waves for constant vorticity flow. We show how the vortex lines, which are straight for a constant vorticity flow, are twisted gently by a passing small-amplitude wave, while the circulation of any closed material curve is still conserved in accordance with the circulation theorem. 

Our results demonstrate that the assumption of constant vorticity in three-dimensional free surface flows is too restrictive to capture the wave dynamics of such systems. Constant vorticity remains a useful assumption only for two-dimensional wave-current systems, and only provided the Laplace equation is satisfied everywhere. Inserting an oscillating line source singularity, for example, as perhaps the simplest and most common model for a wave source, introduces additional vorticity into the flow which is carried downstream as an undulating vorticity sheet at the centre of a critical vortex layer \cite{ellingsen15a}. The resulting flow pattern could not have been foreseen by assuming a constant vorticity field from the outset, essentially the proceedure employed for this system, erroneously from a physical point of view, by Tyvand \& Lepper\o d \cite{tyvand14}. 

\section*{Acknowledgement}
We have benefited greatly from discussions with Professor Peder A.\ Tyvand and suggestions from Professor Christian Kharif.


\end{document}